\documentclass[twocolumn]{aastex61}
\received{January 20, 2017}
\accepted{April 17, 2017}
\submitjournal{ApJL}

\shorttitle{Ram pressure stripping on an elliptical galaxy}
\shortauthors{Sheen et al.}

\begin{document}

\title{Discovery of ram-pressure stripped gas around an elliptical galaxy in Abell 2670}

\correspondingauthor{Yun-Kyeong Sheen}
\email{yksheen@kasi.re.kr}

\author{Yun-Kyeong Sheen}
\affil{Korea Astronomy and Space Science Institute, 776, Daedeokdae-ro, Yuseong-gu, Daejeon, 34055, Korea}

\author{Rory Smith}
\affiliation{Department of Astronomy, Yonsei University, 50 Yonsei-ro, Seodaemun-gu, Seoul, 03722, Korea}

\author{Yara Jaff\'e}
\affiliation{European Southern Observatory, Alonso de Cordova 3107, Vitacura, Casilla 19001, Santiago de Chile, Chile}

\author{Minjin Kim}
\affiliation{Korea Astronomy and Space Science Institute, 776, Daedeokdae-ro, Yuseong-gu, Daejeon, 34055, Korea}

\author{Sukyoung K. Yi}
\affiliation{Department of Astronomy, Yonsei University, 50 Yonsei-ro, Seodaemun-gu, Seoul, 03722, Korea}

\author{Pierre-Alain Duc}
\affiliation{Laboratoire AIM Paris-Saclay, CEA/Irfu/SAp CNRS Universite Paris Diderot, F-91191 Gif-sur-Yvette Cedex, France}

\author{Julie Nantais}
\affiliation{Departamento de Ciencias F\'isicas, Universidad Andres Bello, Fernandez Concha 700, 7591538 Las Condes, Santiago, Chile}

\author{Graeme Candlish}
\affiliation{Instituto de F\'isica y Astronom\'ia, Universidad de Valpara\'iso, Gran Breta\~na 1111, Valpara\'iso, Chile}

\author{Ricardo Demarco}
\affiliation{Departamento de Astronom\'ia, Universidad de Concepci\'on, Casilla 160-C, Concepci\'on, Chile}

\author{Ezequiel Treister}
\affiliation{Instituto de Astrof\'isica, Facultad de F\'isica, Pontificia Universidad Cat\'olica de Chile, Casilla 306, Santiago, Chile}



\begin{abstract}

Studies of cluster galaxies are increasingly finding galaxies with spectacular one-sided tails 
of gas and young stars, suggestive of intense ram-pressure stripping. These so-called ``jellyfish" 
galaxies typically have late-type morphology. In this paper, we present Multi Unit Spectroscopic Explorer 
(MUSE) observations of an 
elliptical galaxy in Abell 2670 with long tails of material visible in the optical spectra, as well as 
blobs with tadpole-like morphology. The spectra in the central part of the galaxy reveals a stellar 
component as well as ionized gas. The stellar component does not have significant rotation, 
while the ionized gas defines a clear star-forming gas disk. We argue, based on deep optical images 
of the galaxy, that the gas was most likely acquired during a past wet merger. It is possible that 
the star-forming blobs are also remnants of the merger. In addition, the direction and kinematics 
of the one-sided ionized tails, combined with the tadpole morphology of the star-forming blobs, 
strongly suggests that the system is undergoing ram pressure from the intracluster medium. 
In summary, this paper presents the discovery of a post-merger elliptical galaxy undergoing 
ram-pressure stripping.

\end{abstract}




\section{Introduction} \label{sec:intro}
A galaxy infalling into a massive cluster halo experiences ram pressure as it
moves through the intracluster medium \citep[ICM;][]{gun72}. The ram pressure 
causes the stripping of gas in galaxies, and sometimes star-forming tails and blobs can 
be associated with the stripped gas around gas-rich disk galaxies 
\citep[e.g.,][]{fum14,ken14}. Late-type galaxies that present these characteristic 
features are often referred to as ``jellyfish'' galaxies. 

Such a jellyfish-like feature has not yet been reported in elliptical galaxies, 
probably due to their gas-poor nature. However, elliptical galaxies are also 
subject to ram pressure when they enter into the ICM. Deep 
X-ray observations have shown peculiar structures of hot gas halos stripped 
from some elliptical galaxies, for instance, NGC 4406 \citep{for79,whi91}, 
NGC 4472 \citep{irw96}, and NGC 4552 \citep{mac06} in the Virgo cluster and 
NGC 1404 \citep{mac05} in the Fornax cluster. Although those elliptical galaxies 
are going through the same mechanism as the star-forming jellyfish galaxies, 
it is unlikely to substantially affect the evolution of the elliptical galaxies due to 
their low levels of star formation \citep{ken99}.

In this Letter, we report an elliptical galaxy that suffers from ram-pressure stripping 
with star-forming blobs and ionized gas tails in Abell 2670 ($z = $ 0.076).
The galaxy was discovered using deep optical images taken in order to search for post-merger
signatures from red sequence galaxies in galaxy clusters \citep{she12,she16}. 
In the deep images the galaxy shows disturbed halo features and is surrounded 
by several blue blobs, which is unusual for an elliptical galaxy (Figure~\ref{fov}).
In order to confirm the association of the blue blobs with the galaxy, we first conducted
IFU (Integral Field Unit) spectroscopic observations of this galaxy using
MUSE (Multi Unit Spectroscopic Explorer) on VLT (Very Large Telescope).
The MUSE data revealed an unexpected complex hidden structure of gas within and 
surrounding the galaxy, which indicates that the galaxy is experiencing strong ram-pressure
stripping in the cluster environment.

\section{Data} \label{sec:obs}

\begin{figure}
\center
\includegraphics[scale=0.38]{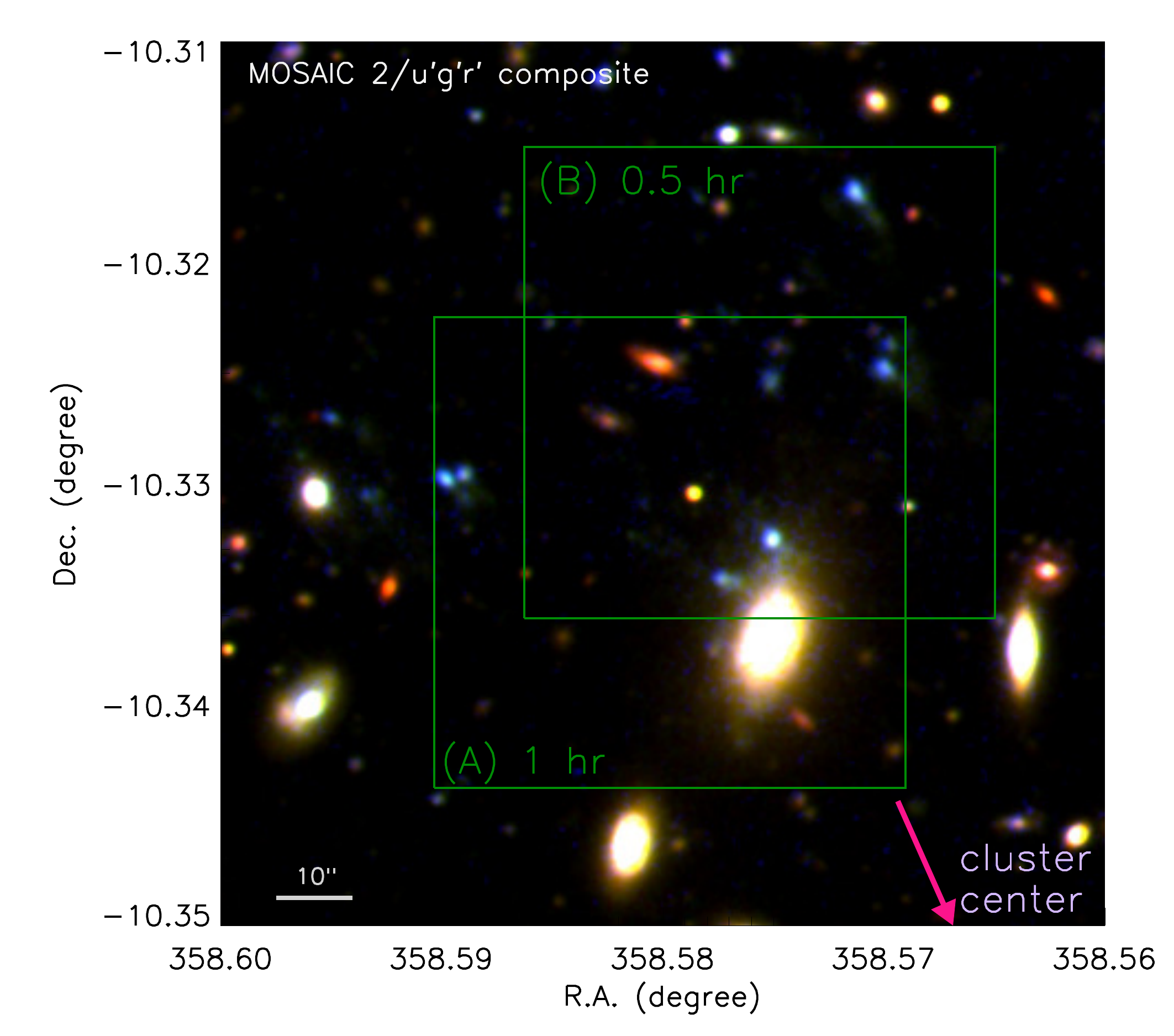}
\caption{MUSE fields of view ($1\arcmin \times 1\arcmin$ for each square) are superimposed 
on a pseudo-color image of the galaxy. The composite image was made of 
$u^{\prime}, g^{\prime}, r^{\prime}$ deep images taken with Blanco/MOSAIC 2.  
(A) and (B) fields were taken with MUSE for 1 hr and 0.5 hr, respectively. Blue blobs were 
discovered in the opposite direction from cluster center. Also, the deep images revealed stellar 
tails of those blue blobs.
\label{fov}}
\end{figure}

We obtained MUSE IFU spectra of the galaxy as a part of ESO program
094.B-0921(A) (PI Yun-Kyeong Sheen). Taking advantage of the revolutionarily 
wide field of view ($1\arcmin \times 1\arcmin$) of MUSE,
it was possible to simultaneously observe the galaxy at $z =$ 0.08 and nearby 
star-forming blobs. In order to include luminous blue blobs at the north of the galaxy, 
we took two MUSE fields of view as shown in Figure~\ref{fov}. The total exposure time 
is 1 hr (4 $\times$ 900 s) and half an hour (6 $\times$ 300 s) for the fields 
(A) and (B), respectively. The MUSE spectral range is 4650 -- 9300\AA~and the spatial 
resolution is 0.${\arcsec}$2 in Wide Field Mode. The data were reduced using MUSE 
pipeline version 1.6 with EsoRex (ESO Recipe Execution Tool) version 3.12.3.
The object frames were stacked to a final data cube for each field. 

\section{Ram-pressure stripping of an elliptical galaxy} \label{sec:rps}
\subsection{Discovery of a gas disk, long ionized gas tails, and star-forming blobs} \label{subsec:tail}

\begin{figure}
\center
\gridline{\fig{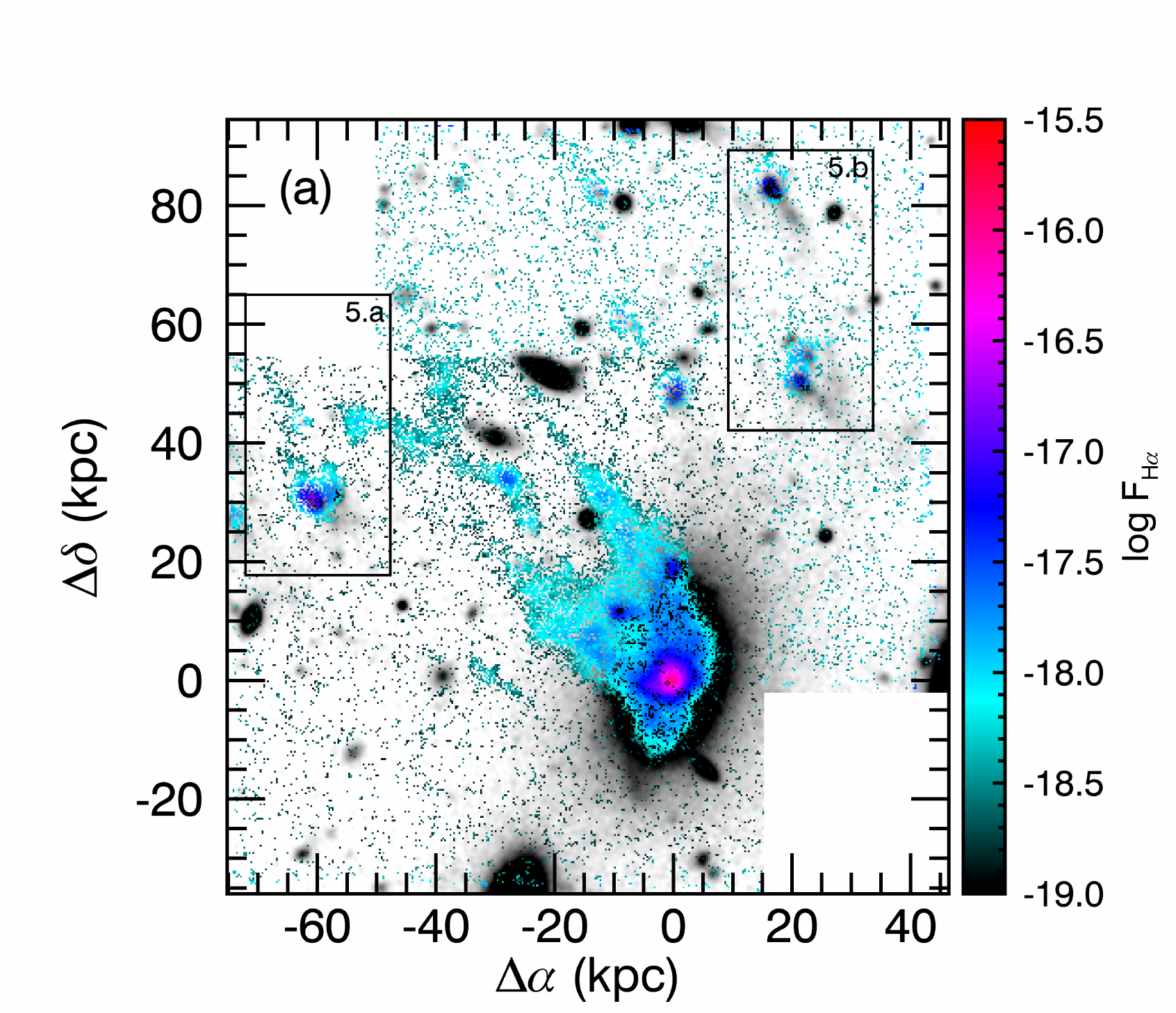}{0.45\textwidth}{}}
\gridline{\fig{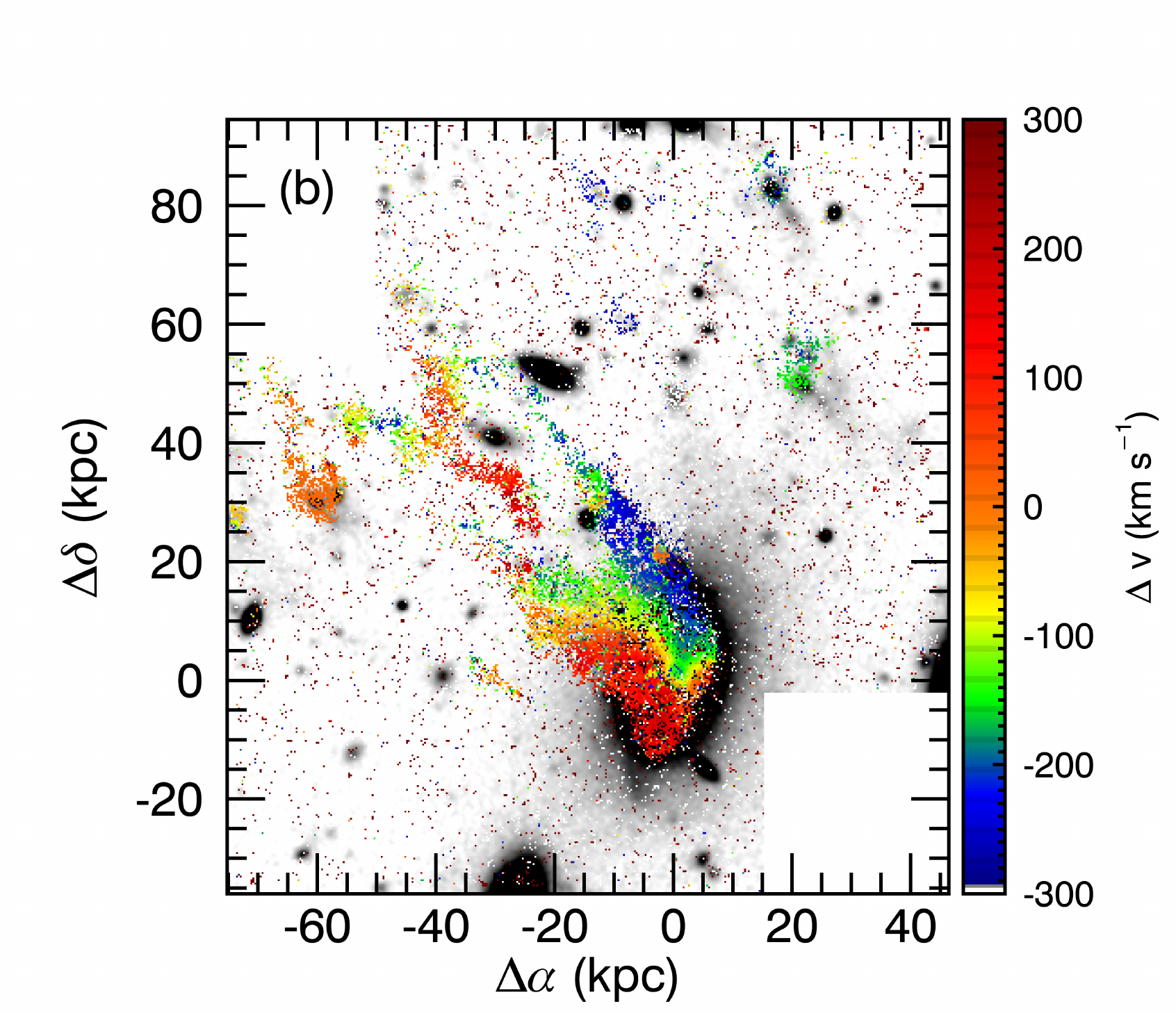}{0.45\textwidth}{}}

\caption{(a) the H$\alpha$ flux map and (b) the velocity offset map 
from the H$\alpha$ of the whole field of view. The zoom-in areas for blue blobs of Figure~\ref{fireball}
are indicated with boxes (5.a and 5.b) in the H$\alpha$ map.  \label{hamap}}
\end{figure}

We performed spectral fitting of the final data cubes using the IDL software KUBEVIZ \citep{fos15}. 
A spectrum from each spaxel was fitted using H$\alpha$ emission lines, as well as continuum spectra,
to derive the H$\alpha$ flux and line of sight velocities with respect to the velocity of the galaxy 
center ($z_{SDSS} =$ 0.08). The entire H$\alpha$ flux map from fields (A) and (B) is presented 
in Figure~\ref{hamap} (a). The MUSE data reveals the presence of an extended H$\alpha$ 
disk at the galaxy center, with bright concentrated emission at the center, and astonishingly 
long ionized gas tails emanating from the disk. The tails are visible out to more than 80 kpc 
from the galaxy. The data also shows H$\alpha$ blobs located at the positions
corresponding to the blue blobs found in our deep optical images (see Figure~\ref{fov}). 

Figure~\ref{hamap} (b) presents a velocity map of the ionized gas using the H$\alpha$ emission line. 
There is a clear indication of rotation in the gas disk of the galaxy, and this velocity field extends into 
the dynamics of the ionized gas tails. The star-forming blobs also seem to well match the velocity field 
of the rotating gas disk and stripped gas. This behavior was also reported in ESO137-001 \citep{fum14}, 
a spiral galaxy experiencing extreme ram-pressure stripping, but this is the first time, to the authors' knowledge, 
that similar behavior has been reported for an elliptical galaxy.

\subsection{Is it really an elliptical galaxy?} \label{subsec:ppxf}
The origin of the gas which is detected in the galaxy is uncertain. From the color 
and morphology, it might be assumed that the galaxy was early-type. If so, then the most 
simple and plausible scenario is that the gas was brought into the galaxy by a recent wet 
merger with a gas-rich companion galaxy. The disturbed stellar halo of the galaxy, revealed 
in the deep optical images, supports this scenario as it shows classical post-merger 
morphological features. However, it is not impossible that the main galaxy might, in fact, 
have been a late-type galaxy, and so the gas was not brought in externally. To try to
understand which scenario is more likely, we studied the morphology and stellar dynamics 
of the galaxy in more detail. 

\begin{figure}
\center
\includegraphics[scale=0.5]{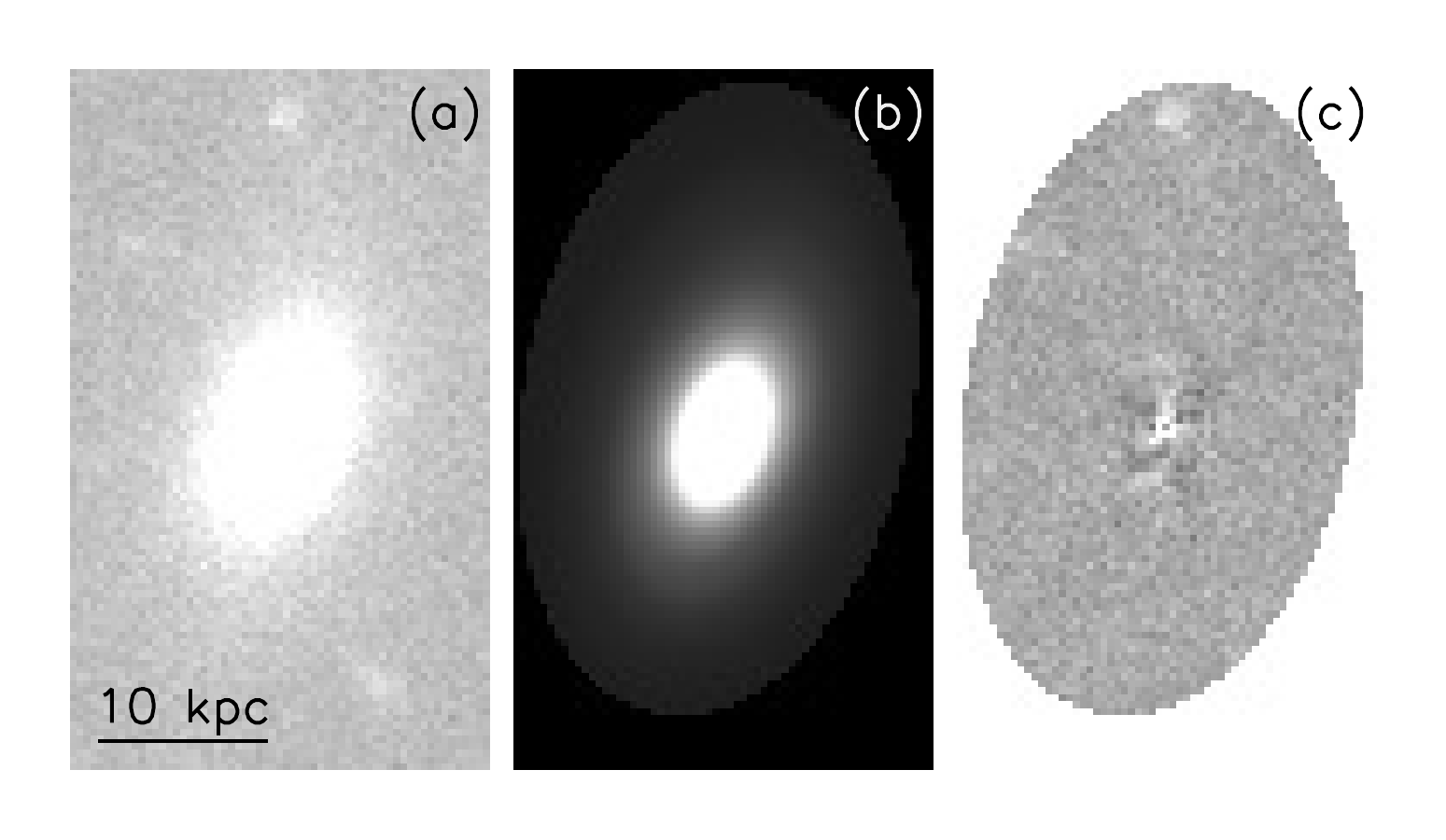}
\includegraphics[scale=0.47]{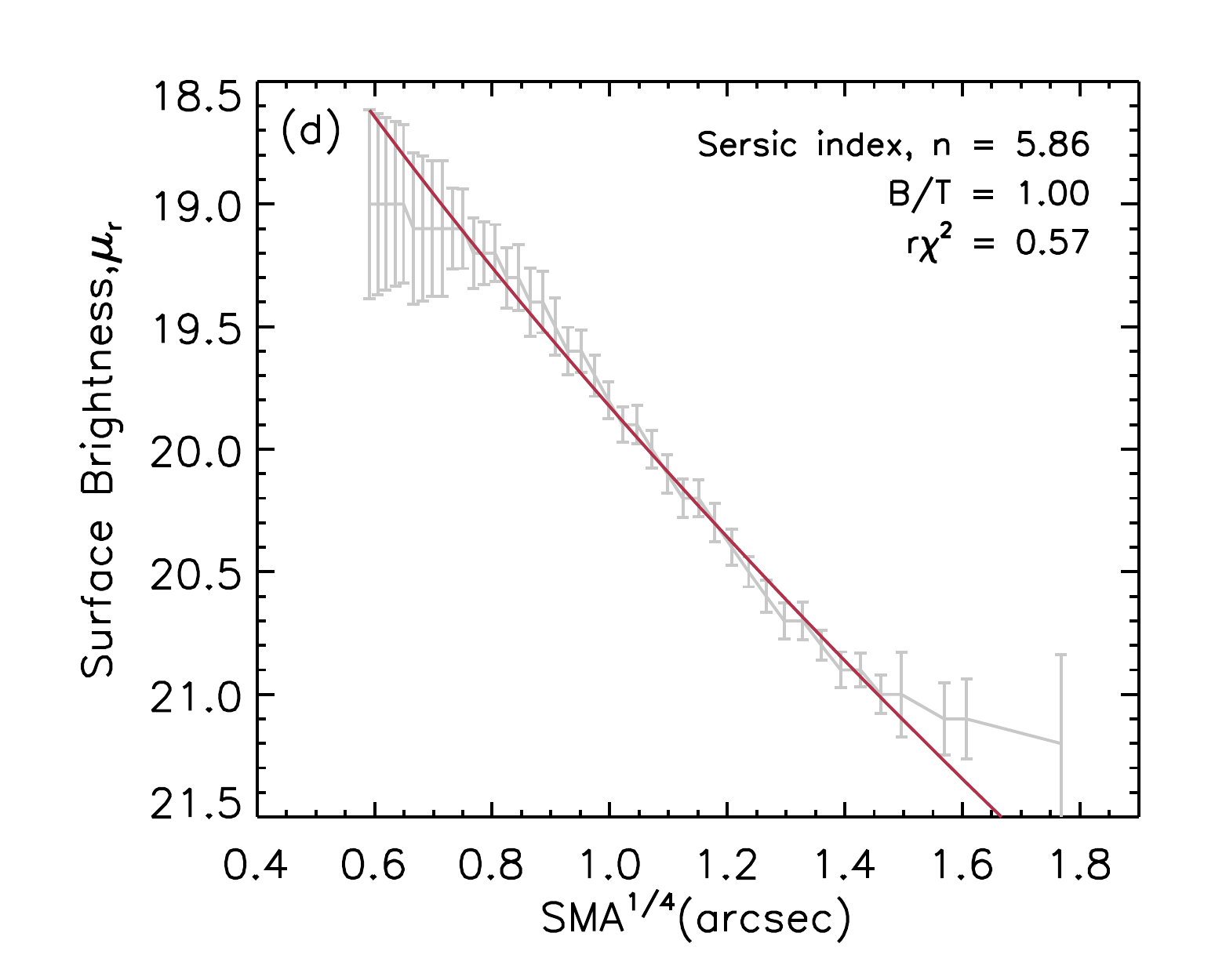}
\caption{(a) MOSAIC 2 $r^{\prime}$-band snapshot image ($t_{exp} = 60~$s), 
(b) \texttt{ellipse} model of the galaxy, (c) a residual image after the model is 
subtracted from the image. The radial surface brightness profile is fitted 
using a combined model of a S\'ersic profile and an exponential profile. The
result is presented in (d). The profile and errors from the data points are presented in gray color
and the best fit is shown by a red line. Although we applied a combined model,
the result suggests that an exponential profile is negligible for this galaxy as B/T = 1. 
The best S\'ersic index, $n$, of the profile also turned out to be greater
than 4 ($n =$ 5.86) using the combined model.
\label{umask}}
\end{figure}

To investigate its morphology, we performed unsharp masking of the galaxy. We used 
an ellipse model, derived by the \texttt{ellipse} task of \texttt{IRAF}. Figure~\ref{umask} 
shows (a) a $r^{\prime}$-band image of the galaxy, (b) a model image
and (c) a residual image after the model was subtracted from the $r^{\prime}$-band image. 
As shown in Figure~\ref{umask} (c), we do not see any hint of a stellar disk in the residual image.
We fitted its radial surface brightness profile using a combined model of a S\'ersic profile 
and an exponential profile, in order to derive a bulge-to-total (B/T) ratio and the best-matching 
S\'ersic index of the galaxy (Figure~\ref{umask} (d)). As a result, we found 
that the B/T ratio $\approx$ 1 and the best-matching S\'ersic index from the combined model 
is greater than 4 ($n =$ 5.86). This result demonstrates that the galaxy is an elliptical galaxy 
with no evidence of a stellar disk.

\begin{figure}
\center
\includegraphics[scale=0.25]{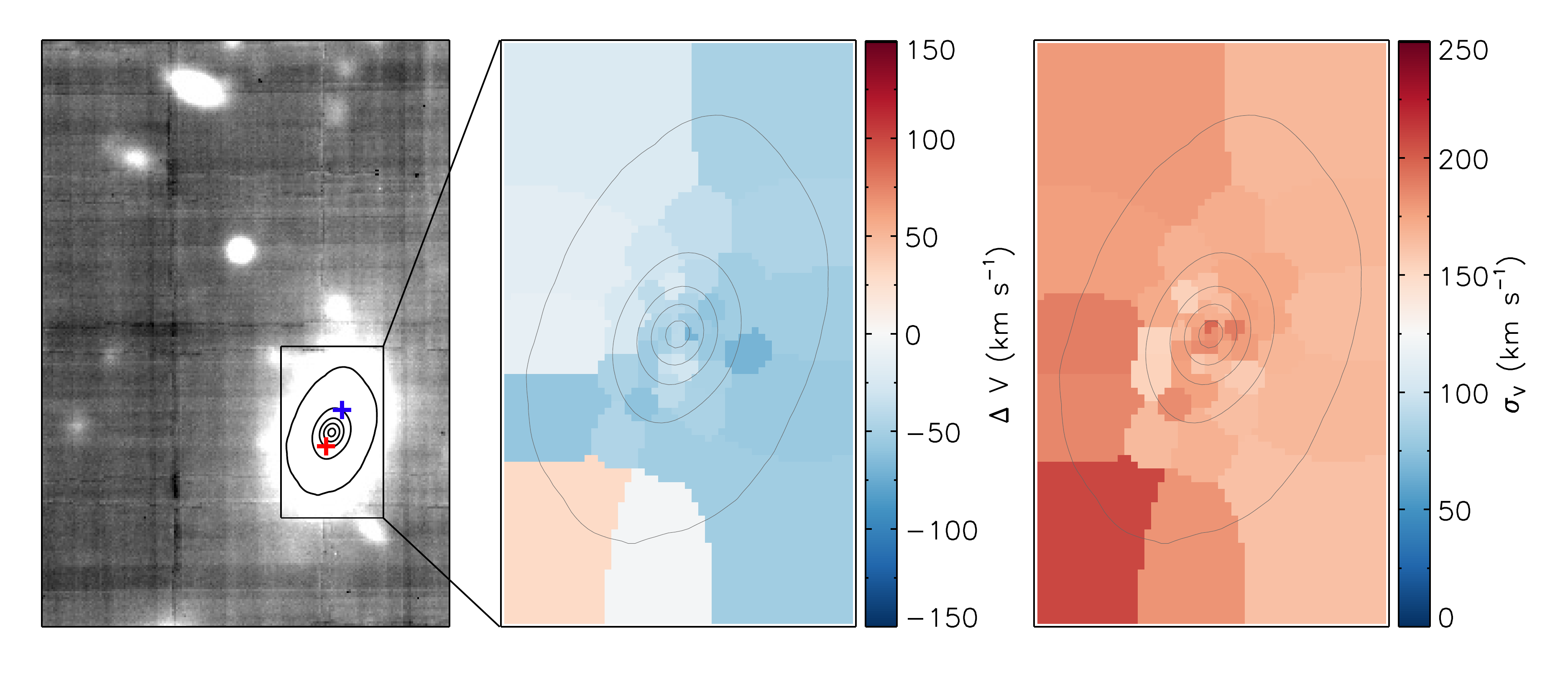}
\includegraphics[scale=0.34]{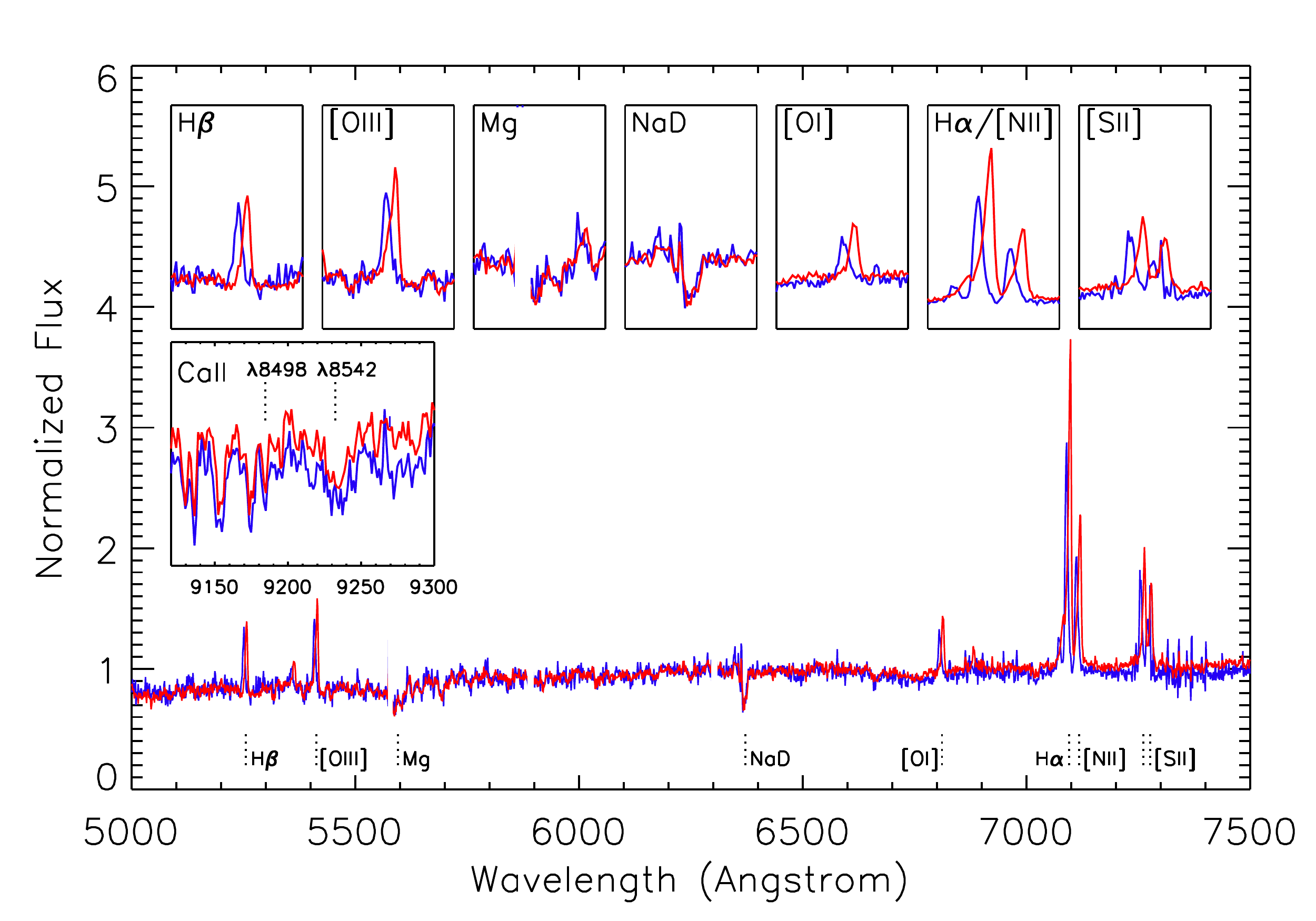}
\caption{Top, from left to right panels: MUSE white-light image, stellar velocity, and velocity dispersion maps derived 
by pPXF. Contours are drawn for different flux levels 
around the galaxy center: 10, 50, 100, 150, and 200 $\times 10^{-20}$ erg s$^{-1}$ cm$^{-2}$ \AA$^{-1}$, 
from the MUSE white-light image. Bottom: MUSE spectra are presented to show the
wavelength shifts of prominent emission lines along the plane of a gas disk, while absorption lines (Mg, NaD, 
and Ca II $\lambda$8498, $\lambda$8542) remain at the same wavelength in the two spectra.
The blue and red spectra correspond to the regions of blue and red crosses
in the white-light image, respectively. The inset images extend for 90 \AA~centered on the line features,
except the two of CaII triplet lines.
The sky residuals are erased in the spectra.  \label{ppxf}}
\end{figure}

We also examined the stellar kinematics of the galaxy using the MUSE data. In the spectra 
of the galaxy's central region, the stellar absorption lines are as prominent as the gas 
emission lines. They indicate that the stellar component consists primarily of old stellar 
populations, while the H$\alpha$ emitting gas disk suggests ongoing star formation. 
We utilized an IDL program, Penalized Pixel-Fitting \citep[pPXF;][]{cap04,cap17}, 
to derive stellar kinematics using absorption lines. The program fitted 
MUSE spectra with the MILES Library of Stellar Spectra \citep{san06}, covering a spectral range of 3525 -- 7500\AA.
Since MUSE spectra start from 4750A, the valid wavelength range for the pPXF was 
4750 -- 8100\AA~(4400 -- 7500\AA~in the rest frame).
Before running the pPXF, the Voronoi 2D-binning method \citep{cap03} was 
applied to the final data cube to achieve S/N $= 30$ at 5500\AA~in each bin.
The velocity offset of stars with 
respect to the galaxy ($z =$ 0.08), and the velocity dispersion in each bin, 
are presented in Figure~\ref{ppxf}. Interestingly, we find that there is no hint of rotation 
in the stars. The galaxy is a dispersion-supported system, with a high velocity dispersion, 
as shown in Figure~\ref{ppxf}. MUSE spectra are presented in the figure
to show the wavelength shifts of prominent emission lines along the plane of a gas disk, while 
absorption lines remain at the same wavelength. Two of Ca II triplet lines
are included in the figure to support that there is no significant rotation of stars, although they were not
utilized for the calculation of stellar velocity offsets.

Based on its surface brightness profile, elliptical shape, and the total lack of rotation
seen in the stellar kinematics, we believe that it is an elliptical galaxy. 
These results are discussed further in Section~\ref{sec:disc}.

\subsection{Morphology of star-forming blobs} \label{sec:streams}

\begin{figure}
\center
\includegraphics[scale=0.75]{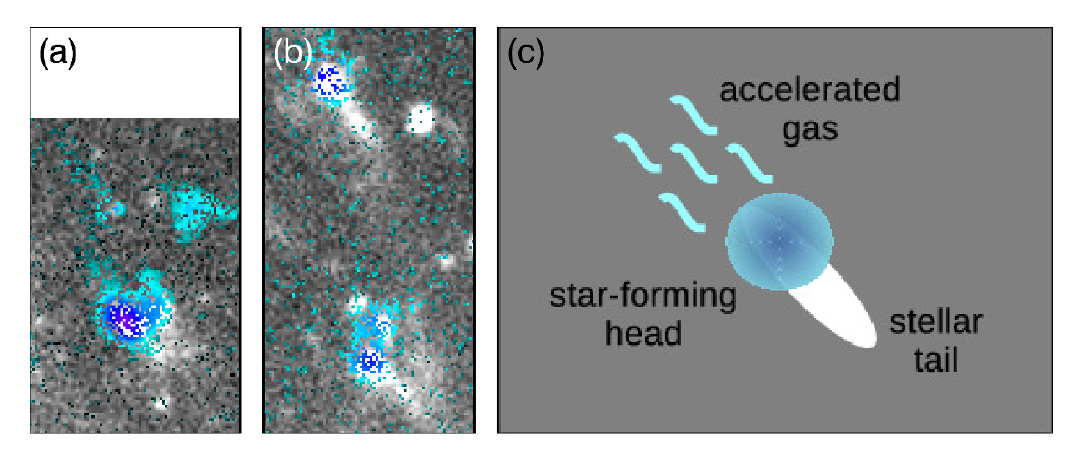}
\caption{(a) and (b) Zoom-in images of some of the star-forming blobs, with the H$\alpha$ flux map
is superimposed over the deep $r^{\prime}$-band image, (c) a schematic view of the blobs. \label{fireball}}
\end{figure}

In Figure~\ref{fireball}, we zoom in on some of the star-forming blobs. 
The H$\alpha$ map is superimposed over the deep $r^{\prime}$-band image. 
We find that the blobs often have a tadpole-like morphology. The head 
of the blobs emits strong H$\alpha$. The blobs also have a white tail, 
visible only in the optical. These are stellar tails, and they tend to point 
toward the cluster center. Previously, objects such as these have been explained 
using the ``Fireball model'' of \citet{ken14}. This suggests that the star-forming 
blobs are experiencing a ram pressure that accelerates them, leaving the stellar 
components behind. For this reason, the stellar tails are expected to point in the 
opposite direction to where the blobs are being accelerated 
(refer to Figure 16 in \citet{ken14}). In that paper, the model is proposed to explain 
the elongated NUV distribution around the ``fireballs'' of the dwarf irregular galaxy 
IC 3418 in the Virgo cluster. However, thanks to  our deep imaging, here the stellar 
tails are revealed clearly in the optical. We also note the presence of streams of 
ionized gas, in the opposite direction to the stellar streams. This is expected if the 
ram pressure, which accelerates the star forming blobs, also strips a little ionized gas 
from the blobs themselves. Thus, under the actions of ram pressure, there is a clear 
segregation, with streams of ionized gas blowing downwind from the blob
and streams of stars pointing upwind. 
A schematic view of the star-forming blobs is shown in Figure~\ref{fireball} (c).
Similar behavior is also expected when tidal 
dwarf galaxies undergo ram pressure stripping (e.g. see Figure 1 of \citet{smi13}).

\section{Discussion} \label{sec:disc}

We report the discovery of an elliptical galaxy with one-sided, long ionized gas tails. 
The tails point away from the cluster center, which suggests that the galaxy is currently 
experiencing ram-pressure stripping in Abell 2670. Simulations of ram-pressure 
stripping have shown that it is easier to strip the more weakly bound ionized hot gas 
from a galaxy than the less-extended cooler disk gas \citep{bek09}. In addition, 
there are several blue star forming blobs, with a tadpole-like morphology, surrounding 
the galaxy. Their stellar tails point upstream, toward the cluster center, further 
supporting the theory that the ionized tail and star forming blobs are actively undergoing 
ram-pressure stripping.

Based on morphology and optical colors, the main galaxy would be classified as  
early-type. The residual image between the galaxy and its ellipse model 
does not show any hint of stellar disk. Moreover, the velocity field of the galaxy, 
measured using the stellar absorption lines, shows no indication of significant 
rotation. Indeed, the velocity dispersions are as large as $\sim$ 200~km s$^{-1}$ 
in the galaxy center. Thus, the stellar kinematics strongly suggest that the galaxy is, 
in fact, an elliptical galaxy, with emanating streams of ionized gas.

Where does the gas come from? A likely scenario is that a wet merger occurred. 
Many merger features are apparent, as indicated by disturbed halo features and 
stellar streams in our deep optical imaging. 

The true origin of the star-forming blobs is uncertain.
In other jellyfish systems, often the star forming blobs external to the galaxy are found 
in a narrow strip, directly downwind of a galaxy that is undergoing ram-pressure stripping. 
This gives the impression that the blobs are forming from the ram-pressure-stripped gas itself 
\citep[e.g. see Figure 1 of][]{ken14}. However, in our case, 
the star-forming blobs are spread over a wide range of angles, beyond the galaxy and its 
ionized gas tails. Therefore, for the blobs to have formed from ram-pressure-stripped gas, 
the ram-pressure wind would have needed to change direction significantly. Moreover, 
the stellar mass range of our star-forming blobs, based on their $g^{\prime} - r^{\prime}$ 
colors \citep{bel03}, is comparable to that of dwarf galaxies 
(log($M_{\star,blob}/M_{\sun}) = 7.6 - 8.4$). Therefore, another possibility is that 
these star-forming blobs are, in fact, tidal dwarf galaxies (TDGs) formed during the wet merger. 
This could explain their large masses and lack of a spatial correlation with the ionized gas stream.

Regardless of whether they formed from ram-pressure-stripped gas, or are TDGs, 
their tadpole-like morphology suggests they are actively experiencing ram pressure now. 
\citet{smi13} demonstrated that ram pressure acting on TDGs produces stellar streams that 
point upwind, in the same way as seen in the ``fireball model'' \citep{ken14}. 
This is because, in both scenarios, a star forming blob of gas is accelerated by ram pressure
and leaves behind a trail of stars, and this occurs independently of how the star forming 
blob was originally formed.

From the H$_\alpha$ emission map, star formation in the center of the galaxy appears very 
centrally concentrated. This, too, is fully consistent with the merger scenario, where gas 
can be driven to a galaxy's center by tidal torques and dissipation, resulting in central starbursts 
\citep{ell13,mor15}. The fact that the main galaxy also shows a very red optical color 
($g^{\prime} - r^{\prime} = 0.99$) implies there must be very heavy internal extinction, 
which reddens the light from the young and blue stellar 
populations recently formed at its center. 
We note that the formation of young stars within the gas disk could result in a small, rotating 
stellar disk in the center. Such an object would resemble kinematically decoupled cores
in early-type galaxies \citep[][i.e., in this case, with a rotating young component 
surrounded by a non-rotating older component]{ems07}. However, so far we do not detect any evidence 
for such a stellar disk. This might suggest that the gas has not been star forming for very long. 
Given the very red optical color of the main galaxy, however, we cannot rule out the possibility 
that strong dust extinction is obscuring our view of such a disk \citep{sti16}. 

To draw stronger conclusions, additional investigation is needed to measure the 
amount of gas in the system, and to get the actual star formation rate. However, this is beyond 
the scope of this discovery paper. We will attempt to answer this, and the many other open questions 
that were raised in this study (e.g., the properties of stellar populations 
in different star-forming regions of the system), in a follow-up paper.

\acknowledgments
S.K.Y. acknowledges support from the Korean National Research Foundation (Doyak program). 
This study was performed under the umbrella of the joint collaboration between Yonsei University 
Observatory and the Korean Astronomy and Space Science Institute.
G.C. acknowledges support from CONICYT PAI No. 79150053.
R.D. gratefully acknowledges the support provided by the BASAL Center for Astrophysics and 
Associated Technologies (CATA). E.T. acknowledges support from CONICYT-Chile grants 
Basal-CATA PFB-06/2007 and FONDECYT Regular 1160999.
This work was co-funded under the Marie Curie Actions of the European Commission (FP7-COFUND).
This work made use of the KUBEVIZ software which is publicly available at 
http://www.mpe.mpg.de/$\sim$dwilman/kubeviz/.

%

\vspace{5mm}
\facilities{VLT:Yepun (MUSE), Blanco (MOSAIC 2 CCD Imager)}

\end{document}